# REPLY TO J.R. Hunter, , M.J.I. Brown  Discussion of Boretti, A., 'Is there any support in the long term tide gauge data to the claims that parts of Sydney will be swamped by rising sea levels?', Coastal Engineering, 64, 161–167, June 2012, Coastal Engineering Volume 75, May 2013, Pages 1–3.

By Alberto Boretti

E-mail a.a.boretti@gmail.com

## ABSTRACT

Hunter and Brown try to demonstrate in their discussion of the long term tide gauge data published in my previous paper that the sea levels are accelerating when they are not. The sea levels are mostly oscillating and certainly not positively accelerating at the present time. As shown in the graphs proposed here after, having an understanding of the oscillatory behaviour of sea levels and by using linear and parabolic fittings but not being selective in the time window to consider, the tide gauge of Sydney exhibits clear multi decadal and inter-annual periodicities but no detectable component of acceleration, similarly to the many others tide gauges of the Pacific or the rest of the world having enough quality and length. If all the long term tide gauges do not exhibit any present accelerating pattern, possibly some simulations and reconstructions may be wrong similarly to the selective assessment of the sea level rise by fitting only the few years of data useful to support the positive acceleration claim.

## THE SYDNEY TIDE GAUGE IS NOT ACCELERATING SIMILARLY TO THE OTHER LONG TERM TIDE GAUGES OF THE WORLD

There are many multi-decadal periodicities of possible influence on the sea levels in the Pacific (NOAA, 2012). The influence of the multi-decadal oscillations on the sea levels and other climate parameters is discussed in Chambers, Merrifield and Nerem (2012); Jevrejava, Moore, Grinsted and Woodworth (2008); Parker, 2012a,b; Mazzarella, Giuliacci and Scafetta, 2012; Scafetta N, 2012a,b,c; Mazzarella and Scafetta, 2012. Quasi-periodic fluctuations of sea levels with a period of about 60 years are explicitly claimed by Chambers, Merrifield and Nerem (2012) and Jevrejava, Moore, Grinsted and Woodworth (2008).

Sydney NSW is the composite record of two tide gauges, SYDNEY, FORT DENISON of time span of data: 1886 – 1993 and completeness (%): 100 and SYDNEY, FORT DENISON 2 of time span of data: 1914 – 2010 and completeness (%): 98. The two records of Sydney are overlapping for almost 80 years with only very minor differences and they can be used to produce a longer record of good quality.

The least squares method is used here to calculate a straight line that best fits the data within a time window and return the window's sea level rate of rise (SLR). The dependent $y$-values are the monthly average sea levels and the independent $x$-values are the time in years. The calculations for $SLR_{j,k}$ is based on the formula:

$$SLR_{j,k} = \frac{\sum_{i=j}^{k}(x_i - \bar{x}) \cdot (y_i - \bar{y})}{\sum_{i=j}^{k}(x_i - \bar{x})^2} \quad (1)$$

In this equation $\bar{x}$ and $\bar{y}$ are the sample means and $j$ and $k$ are the indices of the first and last record of the measured distribution considered for the SLR estimation.

At a certain time $x_k$, $x_j$ is taken as $(x_k-30)$ to compute the $SLR_{30}$, $(x_k-60)$ when computing the $SLR_{60}$, or as $x_1$ when computing the $SLR_A$ over all years of the record. This way, from a measured distribution $x_i, y_i$ for $i=1,N$, it is possible to estimate the time histories of $SLR_{30}$, $SLR_{60}$, and $SLR_A$.

Providing that more than 60–70 years of continuously recorded data, without any quality issues, are available in a given location, the $SLR_{A,k}$ usually returns a reasonable estimation of the velocity of sea level change at the present time $x_k$ and the acceleration may then be computed as

$$SLA_k = \frac{SLR_{A,k} - SLR_{A,k-1}}{x_k - x_{k-1}} \quad (2)$$

This conventional velocity and acceleration might clearly oscillate, and their time history, rather than a single value, is of interest.

In a case with non-accelerating tide gauge records as the norm so far, $SLR_{1,N}$ returns the present sea level rate of rise, and the graphs of $SLR_{j,k}$ and $SLA_k$ are only helpful to confirm the lack of any acceleration. In a case of accelerating tide gauge records as sometimes reconstructed but so far never measured, this approach would confirm the presence of acceleration in the form of a constantly increasing $SLR_{j,k}$ and a constantly positive $SLA_k$ rather than oscillating values about the longer term trend and the zero. At this stage, different mathematics would be needed to compute the present velocity and acceleration.

Figure 1 presents the 12 months moving averages of sea levels, the periodogram of the monthly departures vs. the linear trend and the computed sea level rate of rises (SLR) with 20, 30, 60 years or all the data for Sydney NSW.

The $SLR_A$ is computed only after 20 years of recording. The $SLR_{20}$ and $SLR_{30}$ have large oscillations while the $SLR_{60}$ has these fluctuations significantly reduced. Over the last 60 y, the $SLR_{20}$, $SLR_{30}$ and $SLR_{60}$ have been oscillating without a positive acceleration trend. The present values have been previously recorded and exceeded.

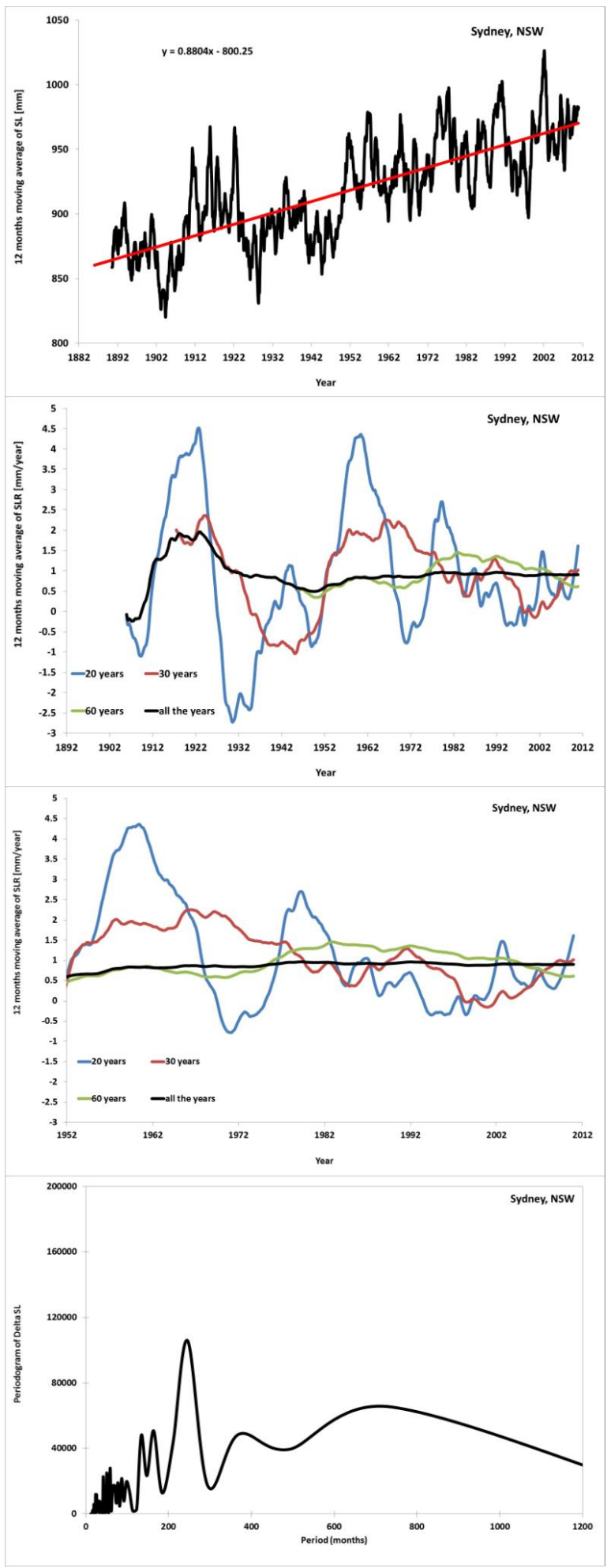

Figure 1 - 12 months moving averages of sea levels, periodogram of monthly departures vs. linear trend and sea level rises (SLR) with 20, 30, 60 years or all the data for the long term tide gauges of Sydney, NSW (data from PSMSL, 2012).

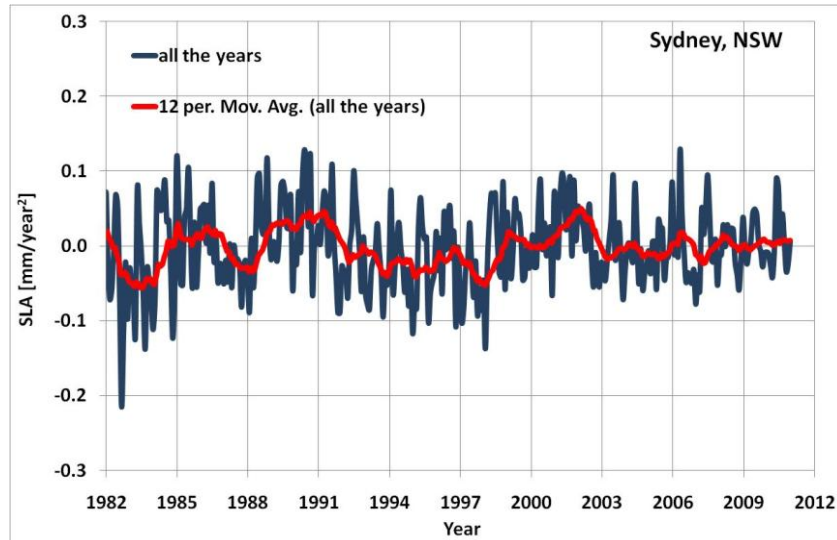

Figure 2 - Sea level acceleration over the last 3 decades for the long term tide gauges of Sydney, NSW (data from PSMSL, 2012).

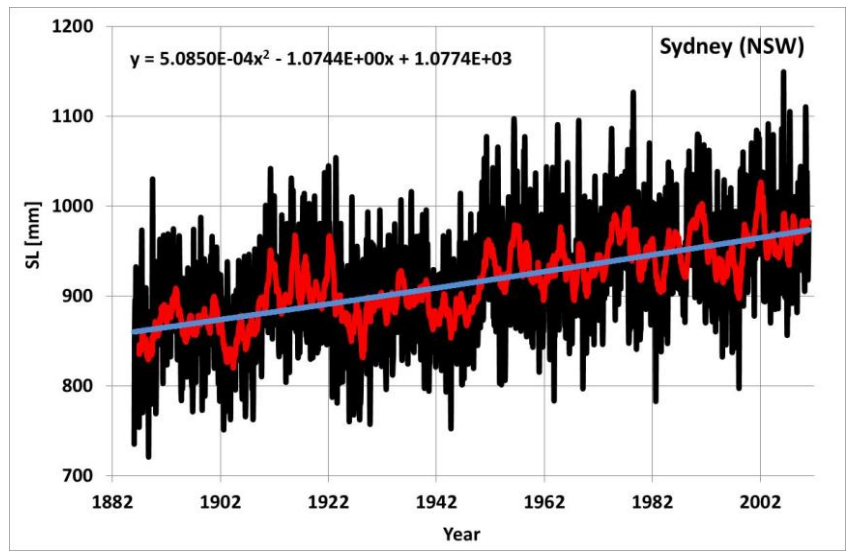

Figure 3 – 2nd order polynomial fittings of the monthly sea levels in the long term tide gauges of Sydney, NSW (data from PSMSL, 2012).

The 12 months moving averages oscillate about the linear trends. The $SLR_A$ approaches the final long-term value after 65-70 years as in Sydney. The present $SLR_{60}$ is not a maximum and the index is regularly oscillating over the last 6 decades.

In case of $SLR_A$ the window length is increasing while for the other curves $SLR_{20}$, $SLR_{30}$ and $SLR_{60}$ the window length is constant. The last point of the $SLR_A$ curve corresponds to the linear fit of the entire record.

Sydney, NSW has very clear multi decadal oscillation of period about 240 months, but also oscillations of period about 400 and above 600 months.

The sea level rate of rise indexes (equation 1) are oscillating around an apparent average value which suggests the existence of natural oscillations. To determine the presence or absence of a present acceleration, a linear fit of the rate curve functions of Figure 1 does not help too much because similarly to the fitting of the original sea level data with a second order polynomial this approach would return an average acceleration over the time of observation and not the present acceleration.

As proposed in Figure 2 for Sydney, NSW, it is graph of the SLA (equation 2) that may show the presence or the absence of acceleration. If the SLA oscillates about zero over the last few decades, then the positive accelerating theory is wrong.

Tables 1 and 2 below present the sea level rate of rise $SLR_A$ (equation 1) and the sea level acceleration (SLA equation 2) computed. The $SLR_A$ is in mm/year and the SLA in mm/year$^2$. Because of the naturally oscillating behaviour of the seas, if the data are updated month by month, the $SLR_A$ and the SLA then oscillates month by month. Averaged values over 10, 20 and 30 years are considered. The computed accelerations are negligible, with small positive and small negative values alternating depending on the window adopted. The differences in between the $SLR_A$ and the magnitudes of the acceleration are well below the limit of accuracy of the measurement and are not further discussed here.

Table 1 – $SLR_A$ averaged over different time windows (values in mm/year).

|  | 10y average | 20y average | 30y average |
|---|---|---|---|
| Sydney, NSW | 0.905 | 0.912 | 0.918 |

Table 2 – SLA averaged over different time windows (values in mm/year$^2$)

|  | 10y average | 20y average | 30y average |
|---|---|---|---|
| Sydney, NSW | 0.004 | -0.004 | -0.003 |

It is worth of mention that only the $SLR_A$ and the SLA graphs in Figure 1 and 2 and the averaged values of these parameters of Table 1 and 2 permit to assess the presence of a positive acceleration at the present time. Houston and Dean (2011) and Watson (2011) both adopted a polynomial fitting of the monthly average mean sea levels to compute the average acceleration of sea levels over the length of the tide gauges as twice the second order coefficient. It has been argued by Donoghue and Parkinson (2011), Rahmstorf and Vermeer (2011) and many others that this average acceleration over the length of the tide gauge record does not tell us if the sea levels are presently accelerating or not. The subject is also covered by Parker A, M. Saad Saleem and M. Lawson (2012) that suggest there are better indicators of the presence or absence of present accelerating patterns. The procedure proposed here is certainly the best option provided so far in the literature to clarify if there is or not a present acceleration.

The $2^{nd}$ order polynomial fitting of the monthly sea levels is presented in Figure 3 only as a reference. The average acceleration over the period of observation is twice the second order coefficients. Sydney has a small positive average acceleration of magnitude of the order of 0.001 mm/year$^2$.

From the analysis of the data measured so far in Sydney, NSW, the sea level is not accelerating here and similar conclusions are offered by all the others tide gauges of the world of enough length and quality if properly analysed.

The lack of positive acceleration in sea levels is consistent with many other analyses indicating an absence of any signs of acceleration (e.g. Boretti and Watson, 2012; Boretti, 2012a,b,c,d,e; Holgate, 2007; Houston and Dean, 2011; Morner, 2004, 2007, 2010a,b; Parker, 2012a,b,c,d,e; Parker, Saad Saleem and Lawson, 2012; Unnikrishnan and Shankar, 2007; Watson, 2011; Wenzel and Schröter, 2010; Wunsch, Ponte and Heimbach, 2007).

In absence of acceleration for all the long term tide gauges of the world, the most likely sea level rise by 2100 (or at least the lower bound of all the catastrophic scenarios) is therefore a mere 78.3 mm in Sydney, NSW. Claims of sea level rises by metres by 2100 are similar to many other claims related to global warming only gross exaggerations.

**REFERENCES**

1) Boretti A and Watson T (2012), The inconvenient truth: Ocean Levels are not accelerating in Australia or over the world. *Energy & Environment* **23**:801-817.

2) Boretti A (2012a), Is there any support in the long term tide gauge data to the claims that parts of Sydney will be swamped by rising sea levels? *Coastal Engineering* **64**:161-167.

3) Boretti A (2012b), Discussion of Natalya N. Warner, Philippe E. Tissot, Storm flooding sensitivity to sea level rise for Galveston Bay, Texas, Ocean Engineering 44(2012); 23-32. *Ocean Engineering*, In Press, Accepted Manuscript, dx.doi.org/10.1016/j.oceaneng.2012.06.030.

4) Boretti A (2012c), Discussion of 'Dynamic system model to predict global sea-level rise and temperature change' by Aral, M.M., Guan, J., Chang, B., Journal of Hydrologic Engineering, Volume 17, Issue 2, 7 March 2012, Pages 237-242. *ASCE's Journal of Hydrologic Engineering*, In Press, Accepted Manuscript, dx.doi.org/10.1061/(ASCE)HE.1943-5584.0000447.

5) Boretti A (2012d), Discussion of Christine C. Shepard, Vera N. Agostini, Ben Gilmer, Tashya Allen, Jeff Stone, William Brooks and Michael W. Beck, Reply: Evaluating alternative future sea-level rise scenarios, Natural Hazards, 2012, DOI: 10.1007/s11069-012-0160-2. *Natural Hazards*, In Press, Accepted Manuscript, dx.doi.org/10.1007/s11069-012-0268-4.

6) Boretti A (2012e), Discussion of J.A.G. Cooper, C. Lemckert, Extreme sea level rise and adaptation options for coastal resort cities: A qualitative assessment from the Gold Coast, Australia, Ocean & Coastal Management, In Press, Accepted Manuscript, Available online 18 April 2012. *Ocean & Coastal Management*, In Press, Accepted Manuscript, dx.doi.org/10.1016/j.ocecoaman.2012.05.031.

7) Chambers, D.P., Merrifield, M.A., Nerem, R.S. (2012), Is there a 60-year oscillation in global mean sea level? *Geophysical Research Letters* **39** (17), art. no. L18607.

8) Donoghue, J.F. and Parkinson, R.W., (2011). Discussion of: Houston, J.R. and Dean, R.G., 2011. Sea-Level Acceleration Based on U.S. Tide Gauges and Extensions of Previous Global-Gauge Analyses. *Journal of Coastal Research*, **27**(3), 409–417

9) Holgate SJ (2007), On the decadal rates of sea level change during the twentieth century. *Geophysical Research Letters* **34**, L01602.

10) Houston JR and Dean RG (2011), Sea-Level Acceleration Based on U.S. Tide Gauges and Extensions of Previous Global-Gauge Analyses. *Journal of Coastal Research* **27**:409–417.

11) Jevrejava S., Moore J.C., Grinsted A., Woodworth P.L. (2008), Recent global sea level acceleration started over 200 years ago? *Geophysical Research Letters* **35** (8), art. no. L08715.

12) Mazzarella A, A Giuliacci and N Scafetta (2012), Quantifying the Multivariate ENSO Index (MEI) coupling to CO2 concentration and to the length of day variations. *Theoretical Applied Climatology*, In Press, Accepted Manuscript, dx.doi.org/10.1007/s00704-012-0696-9.

13) Mazzarella A and N Scafetta (2012), Evidences for a quasi-60-year North Atlantic Oscillation since 1700 and its meaning for global climate change. *Theoretical Applied Climatology* **107**:599-609.

14) Mörner, N.-A., (1995), Earth rotation, ocean circulation and paleoclimate. *GeoJopurnal*, 37:4, 419-430.

15) Mörner, N.-A., (1996), Sea Level Variability. *Z. Geomorphology N.S.*, 102, 223-232.

16) Mörner NA (2004), Estimating future sea level changes. *Global Planetary Change* **40**:49-54.

17) Mörner NA (2007), Sea Level Changes and Tsunamis. Environmental Stress and Migration over the Seas. *Internationales Asienforum* **38**:353-374.

18) Mörner NA (2010a), Some problems in the reconstruction of mean sea level and its changes with time. *Quaternary International* **221**:3-8.

19) Mörner NA (2010b), Sea level changes in Bangladesh new observational facts. *Energy and Environment* **21**:235-249 (2010).

20) National Oceanic and Atmospheric Administration NOAA (2012), Teleconnections. http://www.cpc.ncep.noaa.gov/products/precip/CWlink/daily_ao_index/teleconnections.shtml

21) Parker A (2012a), SEA LEVEL TRENDS AT LOCATIONS OF THE UNITED STATES WITH MORE THAN 100 YEARS OF RECORDING. *Natural Hazards*, In Press, Accepted Manuscript, dx.doi.org/10.1007/s11069-012-0400-5.

22) Parker A (2012b), Oscillations of sea level rise along the Atlantic coast of North America north of Cape Hatteras. *Natural Hazards*, In Press, Accepted Manuscript, dx.doi.org/10.1007/s11069-012-0354-7 (2012).

23) Parker A (2012c), Comment to M Lichter and D Felsenstein, Assessing the costs of sea-level rise and extreme flooding at the local level: A GIS-based approach, Ocean & Coastal Management 59 (2012) 47-62. *Ocean & Coastal Management*, In Press, Accepted Manuscript, dx.doi.org/10.1016/ j.ocecoaman.2012.08.020.

24) Parker A (2012d), Comment to Shepard, C.C., Agostini, V.N., Gilmer, B., Allen, T., Stone, J., Brooks, W., Beck, M.W., Assessing future risk: Quantifying the effects of sea level rise on storm surge risk for the southern shores of Long Island, New York, Natural Hazards, Volume 60, Issue 2, January 2012, Pages 727-745. *Natural Hazards*, In Press, Accepted Manuscript, dx.doi.org/10.1007/s11069-012-0314-2.

25) Parker A, M. Saad Saleem and M. Lawson (2012), Sea-Level Trend Analysis for Coastal Management, *Ocean & Coastal Management*, In Press, Accepted Manuscript, doi: 10.1016/j.ocecoaman.2012.12.005.

26) Scafetta N., (2010). Empirical evidence for a celestial origin of the climate oscillations and its implications. *Journal of Atmospheric and Solar-Terrestrial Physics* **72**, 951-970.

27) Permanent service on Mean Sea Level PSMSL (2012), Sea level data. http://www.psmsl.org

28) Rahmstorf, S. and Vermeer, M. (2011). Discussion of: Houston, J.R. and Dean, R.G., 2011. Sea-Level Acceleration Based on U.S. Tide Gauges and Extensions of Previous Global-Gauge Analyses. *Journal of Coastal Research*, **27**(3), 409–417.

29) Scafetta N (2012a), Does the Sun work as a nuclear fusion amplifier of planetary tidal forcing? A proposal for a physical mechanism based on the mass-luminosity relation. *Journal of Atmospheric and Solar-Terrestrial Physics* **81-82**:27-40.

30) Scafetta N (2012b), Multi-scale harmonic model for solar and climate cyclical variation throughout the Holocene based on Jupiter-Saturn tidal frequencies plus the 11-year solar dynamo cycle. *Journal of Atmospheric and Solar-Terrestrial Physics* **80**:296-311.

31) Scafetta N (2012c), Testing an astronomically based decadal-scale empirical harmonic climate model versus the IPCC (2007) general circulation climate models. *Journal of Atmospheric and Solar-Terrestrial Physics* **80**:124-137.

32) Scafetta N., (2012d). A shared frequency set between the historical mid-latitude aurora records and the global surface temperature. *Journal of Atmospheric and Solar-Terrestrial Physics* **74**, 145-163.

33) Unnikrishnan AS & Shankar D (2007), Are sea-level-rise trends along the coasts of the north Indian Ocean consistent with global estimates? *Global Planetary Change* **57**:301-307.

34) Watson PJ (2011), Is There Evidence Yet of Acceleration in Mean Sea Level Rise around Mainland Australia? *Journal of Coastal Research* **27**:368–377.

35) Wenzel M and Schröter J (2010), Reconstruction of regional mean sea level anomalies from tide gauges using neural networks. *Journal Geophysical Research – Oceans* **115**, C08013.

36) Wunsch R, Ponte R and Heimbach P (2007), Decadal trends in sea level patterns: 1993-2004. *Journal of Climatology* **20**:5889-5911.